\documentstyle[editedvolume,epsf]{crckapb} 

\input psfig.tex

\newcommand{\be}{\begin{equation}}
\newcommand{\ee}{\end{equation}}
\def\lsim{\lower.5ex\hbox{$\; \buildrel < \over \sim \;$}}
\def\gsim{\lower.5ex\hbox{$\; \buildrel > \over \sim \;$}}

\begin{opening}
\title{Adenine Abundance in a Collapsing Molecular Cloud}

\author{Sandip K. Chakrabarti}
\institute{S. N. Bose National Centre For Basic Sciences\\
	JD Block, Salt Lake, Sector-III, Calcutta-700091, India\\
	email: chakraba@boson.bose.res.in}
\author{Sonali Chakrabarti}
\institute{Maharaja Manindra Chandra College\\
20 Ramkanto Bose Street, Calcutta-700003, India\\
and\\
Centre for Space Physics, Salt Lake, Calcutta\\
email:csp@cal3.vsnl.net.in}

\end{opening}

\begin{document}

\begin{abstract}
{\small 
A vital ingredient of DNA molecule named adenine may be produced by successive 
addition of HCN during  molecular cloud collapse and star formation. We compute 
its abundance in a collapsing cloud as a function of the
reaction rate and show that in much of the circumstances the resulting amount
may be sufficient to contaminate planets, comets and meteorites. We introduce a $f$-parameter which 
may be used to study the abundance where radiative association takes place.}
\end{abstract}
\smallskip
{\small

}

\noindent: Accepted for Publication as RAPID COMMUNICATIONS for April 1, 2000, issue of Indian Journal of Physics.
\section{Introduction}

In recent papers, Chakrabarti [1], and Chakrabarti
and Chakrabarti ([2], hereafter referred to as Paper I) explored the possibility of the formation of biomolecules
in star formation region using gas-phase chemistry. Their conclusion was that even in frigid condition in interstellar matter some of the
simplest amino acids such as glycine, alanine etc. could be produced even before the formation of stars and planets.
Paper I also showed that with a choice of reaction rate constant $10^{-10}$, significant adenine may also be produced.
Some preliminary results of amino acids are in [1].

Adenine is a simply produced vital component of the DNA molecule and its significant production may point to an important
clue into the problem of origin of life on planets like ours. Because of this, it is essential to carry out careful analysis on the
reaction rates during the adenine formation. In Paper I, we used the formation of adenine by successive addition of
HCN [3] by using an `average' rate. In the normal circumstances, in gas-phase reaction 
$HCN+HCN \rightarrow H_2C_2N_2$ rate would be small, since they must combine 
by radiative association, i.e., they must radiate a photon when combined together. This is a slow process and
the probability of photon emission could be $1$ in  a few thousand to a few million (T. Millar, private communication).
However as the size of the molecule gets bigger, the process becomes faster. Thus, 
it is likely that for a large enough molecule,
the radiative association may take place at every collision and at this stage, the collisional rate may be used. One possibility
is therefore to assume that after every addition of HCN, the reaction rate goes up by a factor of $f$ ($f$
may be anywhere from $1$ to $100$ or more). Hence one may imagine that at the early stages, $HCN+HCN\rightarrow H_2C_2N_2$
forms with a reaction rate of $10^{-16}$, but for $HCN+H_2C_2N_2$ the rate becomes $f\times 10^{-16}$, for $HCN+H_3C_3N_3$ 
the rate becomes $f^2 \times 10^{-16}$ and so on. It would be therefore interest to learn, whether significant adenine 
is formed and it is detectable when the radiative association process is taken into account. In the 
present paper, we do just that. It is possible that more favorable reactions take place on ice, but in view of 
little known  reaction rates of ice chemistry we believe that the best we could do is to study 
the formation of these important molecules as a function of two parameters, namely, $\alpha_{Ad}$ 
and $f$. It is quite possible that a suitable $f$ parameter
we suggested above would take care of the ice-chemistry reaction rates as well.  It is still possible that such an $f$
may actually be determined by actual detection of molecules in space. Similarly, constancy of $f$ is an assumption of our model.
In reality it could vary with the size of the molecules.

So far, there has been controversy whether glycine has been observed in interstellar matter. Miao et al. [4] tentatively
detected glycine in the massive star forming region Sgr B2(N) though this was later challenged by Combes et al. [5]
who suggested that with the sensitivity of the detector taken into account, 
the lines were really at the confusion limit and positive identification would
require more sensitive instruments. It is not known if any attempts were made to detect
adenine lines, however there may have been detection of adenine in 
meteoritic samples (M. Bernstein, private communication).

\section{Reaction Network and Hydrodynamic Model}

We choose the same reaction network as in Paper I which we again present here 
for the sake of completeness.  We take the UMIST database (Millar, Farquhar \& Willacy [6]) as our basis of chemical
reactants and reactions, but added several new reactions such as synthesis of amino acids
(alanine and glycine), hydroxy-acids (glycolic and lactic acids),
DNA base (adenine, [3]), urea synthesis etc.
These new reactions make the total number of species to be $422$.
The rate co-efficients of these additional reactions are difficult to find,
especially in the environs of a molecular cloud. To use UMIST database, the rate
constant for a two body reaction is written as [6],
$$
k=\alpha (T/300)^\beta {\rm exp}(-\gamma/T) \ \ {\rm cm}^3\ s^{-1}
\eqno{(1)}
$$
where, $\alpha$, $\beta$ and $\gamma$ are constants and $T$ is the temperature.
Amino acid synthesis rate was estimated from Fig. 8 of Schulte \& Shock [7].
Urea synthesis rate is kept comparable to the rates given
in UMIST table. The rate constants were taken to be $\alpha=10^{-10}$,
$\beta=\gamma=0$ for each two-body reactions. In Paper I, the rate constants for adenine
synthesis was chosen to be similar to other two body reactions [$\alpha_{Ad}=
10^{-10}$ $\beta=\gamma=0$ for each HCN addition in the chain
$HCN \rightarrow (\alpha_{1})\rightarrow CH(NH)CN \rightarrow (\alpha_{2})\rightarrow NH_2CH(CN)_2 \rightarrow  (\alpha_{3})\rightarrow 
NH_2(CN)C=C(CN)NH_2 \rightarrow (\alpha_{4})\rightarrow H_5C_5N_5$ (adenine)]. In the present paper
we run the same simulation with $\alpha_{i}|_{i=1,2,3,4}=\alpha_{Ad}=10^{-12}, \ 10^{-14},\ 10^{-16}$
as well and in addition, consider the possibility that $\alpha_{i}=f^{i-1}\alpha_{Ad}$ as discussed 
in the introduction. In this notation, Paper I, represents the case with $f=1$.

Initial composition of the cloud before the
simulation begins is kept to be the same as in [6], and formation
of $H_2$ is included using the grain-surface reaction with rates as in [6].
The initial {\it mass fractions} are taken to be the same as in [6]
(but appropriate convertion), i.e., H:He:C:N:O:Na:Mg:Si:P:S:Cl:Fe = $0.64$:$0.35897$:$5.6\times
10^{-4}$:$1.9\times 10^{-4}$:$1.81\times 10^{-3}$:$2.96\times 10^{-8}$:
$4.63\times 10^{-8}$:$5.4 \times 10^{-8}$:$5.79\times10^{-8}$:$4.12\times
10^{-7}$:$9\times 10^{-8}$:$1.08\times10^{-8}$.

The hydrodynamic model is kept same as that in Paper I. We choose the initial
size of the molecular cloud to be $r_0=3\times 10^{18}$cm,
average temperature of the cloud $T=10$K, and angular velocity of the cloud $\Omega=10^{-16}$ rad s$^{-1}$.
In this case the speed of sound is $a_s \sim 19200$cm s$^{-1}$
and corresponding initial density [8]
is $\rho=10^{-22}$g cm$^{-3}$ and accretion rate is ${\dot M}=1.06 \times 10^{20}$g s$^{-1}$. 
In the isothermal phase of the cloud collapse, density $\rho \propto r^{-2}$ [9]
and the velocity is constant. When opacity becomes high enough to trap radiations 
(say, at $r=r_{tr}$), the cloud collapses adiabatically with $\rho \propto r^{-3/2}$. In presence of rotation,
centrifugal barrier forms at $r=r_c$, where centrifugal force balances gravity.
Density falls off as $\rho \propto r^{-1/2}$ in this region [10]. 
The initial constant velocity of infall becomes $8900$cm s$^{-1}$ and below $r=r_c$
velocity $\propto r^{-1/2}$ was chosen to preserve the accretion rate in a disk like structure of constant height.
Since for the parameters chosen (generic as they are) $r_c>r_{tr}$,
we chose $T\propto 1/r$ inside the centrifugal barrier ($r<r_c$) as in an adiabatic flow. We follow the
collapse till a radius of $10^{12}$cm is reached.

\section{Models and Results}

We use the following models parameterized by $\alpha_{Ad}$ and $f$.

\noindent Model A: $\alpha_{Ad}=1.e-16$ and $f=1$.\\
\noindent Model B: $\alpha_{Ad}=1.e-14$ and $f=1$.\\
\noindent Model C: $\alpha_{Ad}=1.e-12$ and $f=1$.\\
\noindent Model D: $\alpha_{Ad}=1.e-10$ and $f=1$ (same as in Paper I)\\
\noindent Model E: $\alpha_{Ad}=1.e-16$ and $f=100$.\\
\noindent Model F: $\alpha_{Ad}=1.e-14$ and $f=10$.\\
\noindent Model G: $\alpha_{Ad}=1.e-12$ and $f=5$.\\

Fig. 1 shows the evolution of adenine abundance $X_{Ad}$ with  time (upper axis in seconds) and with
logarithmic radial distance (in cm). We note that generally speaking, at $r=10^{16}$cm, 
already the abundance $X_{Ad}$ has reached almost the saturated value. This is because, as the
upper axis indicates, most of the time is spent in this region during collapse. The final abundance 
in different models are: (A) $X_{Ad}=1.36\times 10^{-34}$, (B) $1.36\times 10^{-26}$, (C) $1.36\times 10^{-18}$, 
(D) $6.35\times 10^{-11}$, (E) $1.2\times 10^{-22}$, (F) $1.34\times 10^{-20}$, and (G) $1.8\times 10^{-14}$ respectively. 
Note that when $X_{Ad}$ is really small, it is proportional to $\alpha_{Ad}^4$ (for fixed $f$) as expected from a 
reaction with four sequence (see, e.g, Models A, B and C). But when its abundance it significant, HCN otherwise
participating in other reactions also 
contribute significantly to adenine formation. Similarly, at low $X_{Ad}$, for $f\ne 1$, the final abundance is  proportional to
$f^6$ for the same value of $\alpha_{Ad}$.

If the present detectability limit of abundance is around $10^{-11}$ [5], then it is clear that 
adenine processed in our method should not be detectable (except for Model D) 
even though it may be enough to contaminate and flourish in some planets as we suggest. 
With a molecular weight of $135$ for adenine, one could imagine that an abundance of $10^{-21}$ or less
should really be considered as insignificant as far as the contamination theory goes. In that case Models A, B  and E
must be rejected. This would correspond to a lower limit of $<\alpha_{Ad}>$ as $\sim 10^{-13}$ 
(where we use $<>$ to indicate an average over the whole chain of reactions leading the adenine 
formation from $HCN$.). On the other hand, even when $\alpha_{1} \sim <\alpha_{Ad}>$, $f$ could be large enough to 
have eventual significant production (Model F). Thus, our $\alpha-f$ model implies that 
one needs to study the reaction rate of not only $HCN+HCN \rightarrow H_2C_2N_2$, 
but also every stages of HCN addition in order to come a definitive  conclusion in this regard.

\section{Conclusion}

In presence of radiative association, adenine abundance $X_{Ad}$ in an interstellar cloud seems to be roughly proportional to
$\alpha_{Ad}^4 f^6$ for small $X_{Ad}$. This means that the measurements of both $\alpha_{Ad}$ and $f$ must be made very accurately.
We studied the $\alpha-f$ parameter space and found that while some region could produce significant abundance, a smaller region
produce detectable (with present day technology) amount, while the rest produces abundance insignificant enough to
dismiss the contamination theory. One must wait for the technological advancements to improve
laboratory experiments in extreme conditions and to  improve the detectability limit in order to come to a firm conclusion.

\vspace{0.3cm}

\noindent {\bf Acknowledgments}\\

\noindent SC acknowledges the usages of the facilities of Centre for Space Physics for writing this article.

{}

\newpage

{\bf Figure 1.}: Evolution of adenine abundance with radial distance and time (upper axis) in a $\alpha-f$ model.
Various models are marked on the curve. See text for details. 

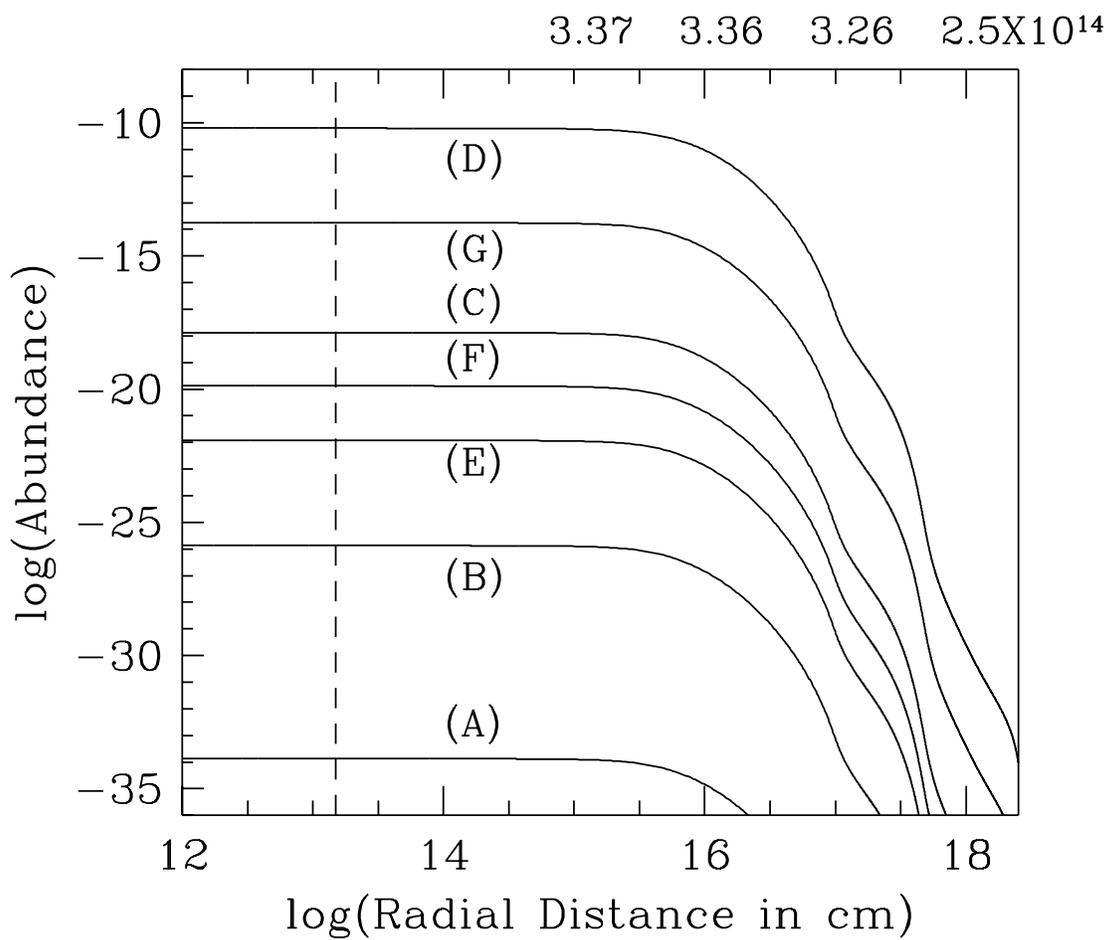
\end{document}